\documentclass{elsart}
\usepackage{graphicx}
\usepackage{amssymb}

\begin{document}

\begin{frontmatter}

\title{Quantum Correction of Fluctuation Theorem}

\author{T. Monnai}
\ead{monnai@suou.waseda.jp}
, \author{S. Tasaki}
\address{Department of Applied Physics ,Waseda University,3-4-1 Okubo,
Shinjuku-ku,Tokyo 169-8555,Japan }

\begin{abstract}
Quantum analogues of the transient and steady-state fluctuation theorems 
are investigated for a harmonic oscillator linearly coupled with a 
harmonic reservoir. The probability distribution for the 
work done externally is derived and the following facts are shown:
(i) \ In the transient fluctuation theorem, there appears 
a quantum correction of order $\hbar^2$.
(ii) \ In the steady-state fluctuation theorem, the existence 
of a quantum correction depends on the way of driving. In the 
uniformly dragged case, the classical formula holds, while, in the
periodically driven case, there appears a correction of order
$\hbar^2$.
\end{abstract}

\end{frontmatter}

\section{INTRODUCTION}

In the linear nonequilibrium regime, statistical properties of fluctuations
are known to characterize nonequlibrium states via exact identities
such as the Green-Kubo formula and Fluctuation-Dissipation Theorems.
On the contrary, general properties of the far-from-equilibrium fluctuations
have not been well understood. Recently, two new identities valid even far
from equilibrium were found: the fluctuation theorem (FT) first found by
Evans, Cohen and Morriss\cite{EvansCohen} and the nonequilibrium
free-energy equality given by Jarzynski\cite{JarzynskiEq}.

The fluctuation theorem addresses the symmetry of entropy production or work
externally done and several different versions are known. Nevertheless, it
takes the following general form:
\begin{equation}
\left(\lim_{\tau\to +\infty} \right){1\over \tau}\log 
\frac{{\rm Prob}(\sigma)}{{\rm Prob}(-\sigma)} = \sigma \ ,
\label{FT}
\end{equation}
where ${\rm Prob}(\sigma)$ stands for the probability of observing
an average entropy production rate (or an average power applied externally),
$\sigma$, during the time interval $\tau$. The $\tau\to +\infty$ limit may or may not
be taken depending on the situations. 

FT with $\tau\to +\infty$ limit was first found numerically by Evans, Cohen and 
Morriss\cite{EvansCohen} for nonequilibrium steady states of thermostatted systems. 
Then, it was rigorously proved by Gallavotti and Cohen\cite{Gallavotti1}. This version 
is referred to as the steady-state fluctuation theorem (SSFT). On the other hand, before 
the proof by Gallavotti and Cohen, Evans and Searles\cite{EvansSearles0} derived the 
relation (\ref{FT}) without $\tau\to +\infty$ limit for transient trajectories starting 
from initial states obeying the microcanonical distribution. This version is refered to
as the transient fluctuation theorem (TFT) and has been extensively studied by Evans 
et al.\cite{Evans}. 
Moreover, it was extended to stochastically driven systems\cite{KurchanCl,Spohn1999,Maes} 
and to open conservative systems\cite{Jarzynski,RondoniTel,ReyBellet}. Its relation with 
Jarzynski equality was studied as well\cite{12} and the related topics have been 
extensively investigated (see e.g., references in Refs.\cite{ReyBellet,Maes2}).

Recently, a beautiful experiment by Wang et al.\cite{1} confirmed TFT for a colloidal
particle kept in a uniformly moving optical trap. The dynamics of the colloidal particle
is governed by the Langevin equation and TFT holds for the work done externally. 
FT for the Langevin equation was studied by Mazonka and Jarzynski\cite{7} to illustrate 
the difference between SSFT and TFT, and the experiment stimulates the reinvestigations 
of FT for the Langevin equations\cite{8,6,vanZon,Dhar}.

Contrary to FT for classical systems, FT for quantum systems has not
been well understood. In order to generalize FT to quantum systems, it is
necessary to identify entropy change or work done externally. And two procedures
are available:

\begin{itemize}

\item[(a)] Measure energy, particle numbers etc. twice and evaluate
the entropy change or work done as the difference of the two observed 
values.

\item[(b)] Measure flows of energy, particle numbers etc. and evaluate
the entropy change or work done as accumulated values of the flows.

\end{itemize}

\noindent
Needless to say, in classical systems, the two procedures give the same 
values of the entropy change or work done, because of the conservation 
of energy and particle numbers.  However, in quantum systems, as a result 
of the noncommutativity of dynamical variables, the two procedures
are not equivalent. 

As far as the authors know, the procedure (a) was first suggested by
Kurchan\cite{9} and he showed that the fluctuation theorem (\ref{FT})
holds for the probability distribution function of the so-obtained 
entropy change or work done. 
Similar results were discussed by Hal Tasaki\cite{HalTasaki},
Callens et al.\cite{Callens}, and the C$^*$ generalization was given by 
Tasaki and Matsui\cite{TasakiMatsui}. 

On the other hand, the procedure (b) has not been 
well studied. In this paper, we investigate a quantum analogue of 
the fluctuation theorem, following the procedure (b), with respect to 
the work done externally for a harmonic oscillator linearly coupled with 
a harmonic reservoir. The oscillator is externally driven in such a way that 
the center of the harmonic potential follows a given trajectory. 
Note that the system is a quantum analogue of the Langevin equation\cite{7,8,6,vanZon,Dhar} 
which describes the experiment by Wang et al.\cite{1}.
Then, we have found that (i) \ there appears a quantum correction of
order $\hbar^2$ in TFT, and that (ii) 
\ the existence of a quantum correction for SSFT 
depends on the way of driving (no correction
in the uniformly dragged case and $\hbar^2$-correction in the
periodically driven case). Moreover, the quantum correction for
SSFT is universal in the sense that
it depends only on the temperature and the frequency of the driving, 
but not on the specific features of the system and environment.

The rest of this paper is arranged as follows. The model is described 
in Sec.\ref{Sec.Model}. The equation of motion is solved and quantum
analogues to the fluctuation theorems are derived in Sec.\ref{Sec.Dis}.
The subsequent sections are devoted to the discussions of TFT
and SSFT separately.
In the last section, discussions are given.

\section{MODEL}\label{Sec.Model}

In order to discuss quantum corrections to the fluctuation theorem,
we study an exactly solvable model of a one-dimensional harmonic 
oscillator linearly coupled with a harmonic 
bath\cite{Ford,10}.
Because a straightforward derivation of a quantum Langevin equation
is known\cite{10}, the Gardiner's version is used.
Then, we consider the following thought experiment:

\begin{itemize}

\item[(i)] The system is prepared to be in equilibrium with inverse 
temperature $\beta$\footnote{Throughout this article, we use the unit 
where the Boltzmann constant is unity.}.

\item[(ii)] At time $t=0$, the harmonic potential starts to move, 
where its center follows the trajectory $x=f(t)$. 

\end{itemize}

\noindent
When $t<0$, the potential center is assumed to be fixed at $x=0$ and 
to move continuously (namely, $f(0)=0$).
The Hamiltonian which describes the time evolution is given as 
\begin{eqnarray}
H(t)&=& H_0 + \frac{k}{2} f(t)\left(-2 q+f(t)\right) \ , \\
H_0&=&\frac{p^2}{2m}+\frac{k}{2} q^2+
\frac{1}{2}:\int d\lambda((p_\lambda-\kappa_\lambda q)^2
+\omega_\lambda^2 q_\lambda^2): \ , \label{Hamiltonian0}
\end{eqnarray}
where $q$, $p$, and $m$ stand for the coordinate, momentum and mass 
of the harmonic oscillator, respectively, $k$ is the strength of the 
harmonic potential, $q_\lambda$ and $p_\lambda$ are the coordinate and
momentum of a bath degree of freedom with frequency $\omega_\lambda$,
and $:\cdots :$ is the normal product with respect to the normal modes:
\begin{equation}
a_\lambda\equiv \frac{1}{\sqrt{2\hbar \omega_\lambda}} (\omega_\lambda 
q_\lambda+i p_\lambda) \ . \label{OldMode}
\end{equation}
Here we assume that $\omega_\lambda$ runs from 0 to $+\infty$ and that
the dispersion equation
\begin{equation}
\eta(z)\equiv mz^2-k-\int d\lambda \kappa_\lambda^2 - \int d\lambda {\omega_\lambda v_\lambda^2
\over z^2-\omega_\lambda^2} \label{DisEq} \ 
\end{equation}
has no real zeros, where 
\begin{equation}
v_\lambda\equiv \sqrt{\omega_\lambda}\kappa_\lambda \ . \label{intV}
\end{equation}
This condition corresponds to the case of the damped harmonic oscillator.

As in the previous works\cite{1,8,6,7}, 
we investigate the statistical property of the work externally done
during the time interval $\tau$ divided by temperature $\beta^{-1}$:
${\hat \Sigma}_\tau$. 
If the work is started to be measured at time $t$, it is given by
\begin{equation}
{\hat \Sigma}_\tau \equiv \beta \int_0^\tau \dot{f}(s+t)(-k(q(s+t)-f(s+t)))ds
\end{equation}
where $\dot f$ stands for the time derivative of $f$ and
$q(s+t)$ is the coordinate operator in the Heisenberg picture at time $s+t$. 
Then, the probability density 
$\pi_t(\Sigma_\tau=A)$ of the work is given by
\begin{equation}
\pi_t(\Sigma_\tau=A) = \left\langle e^{-\beta H_0}\delta(
{\hat \Sigma}_\tau-A)\right\rangle
\end{equation}
where $\langle \cdots \rangle \equiv {\rm tr}(e^{-\beta H_0}\cdots )/Z$ stands for the 
thermal average and $Z={\rm tr}e^{-\beta H_0}$ is the partition function.
Note that TFT refers to the symmetry of 
$\pi_0(\Sigma_\tau=A)$ while SSFT refers to
that of $\pi_{+\infty}(\Sigma_\tau=A)$.

\section{DISTRIBUTION FUNCTION OF THE WORK}\label{Sec.Dis}

In this section, we derive an explicit expression of the distribution
function $\pi_t$ of the work $\Sigma_\tau$.

\subsection{Evaluation of ${\hat \Sigma}_\tau$}

First we note that the unperturbed Hamiltonian $H_0$ is diagonalized by
the normal modes:
\begin{equation}
\mskip -30mu 
\alpha_\lambda=a_\lambda+\frac{v_\lambda}{\eta_+(\omega_\lambda)}
\left\{{p-i m \omega_\lambda q\over \sqrt{2\hbar}}+ \int 
{d{\lambda '}\over 2}\left(\frac{v_{\lambda '} a_{\lambda '}
}{\omega_{\lambda }-\omega_{\lambda '}+i 0}
-\frac{v_{\lambda '}a_{\lambda '}^+}{\omega_{\lambda }+\omega_{\lambda '}}\right)\right\}
\label{NewMode}
\end{equation}
where the normal mode $a_\lambda$ and the interaction strength $v_\lambda$
are given, respectively, by (\ref{OldMode}) and (\ref{intV}),
$\eta_+(\omega_\lambda)\equiv \eta(\omega_\lambda+i0)$ and the function $\eta(z)$ is
defined in (\ref{DisEq}).
Indeed, one has $[\alpha_\lambda,\alpha_{\lambda '}^+]=\delta(\lambda-\lambda ')$ and
\begin{equation}
H_0=\int d\lambda \hbar \omega_\lambda \alpha_\lambda^+ \alpha_\lambda +({\rm c-number}) \ .
\end{equation}
The old variables can be represented by these normal modes. For example,
\begin{equation}
q=\sqrt{\frac{\hbar}{2}}\int d\lambda\left(\frac{i v_\lambda}{\eta_-(\omega_\lambda)}
\alpha_\lambda-\frac{i v_\lambda}{\eta_+(\omega_\lambda)}\alpha_\lambda^+\right)
\end{equation}
and, thus, 
\begin{equation}
H(t)=H_0- \sqrt{\frac{\hbar}{2}} k f(t) \int d\lambda \left\{\frac{i v_\lambda \alpha_\lambda}
{\eta_-(\omega_\lambda)}
-\frac{i v_\lambda \alpha_\lambda^+}{\eta_+(\omega_\lambda)}\right\}
+({\rm c-number}) \label{NewHam}
\end{equation}
where $\eta_-(\omega_\lambda)=\eta(\omega_\lambda-i0)$. 

From (\ref{NewHam}), one finds that the normal mode $\alpha_\lambda(t)$ in the
Heisenberg picture obeys
\begin{eqnarray}
\dot{\alpha}_\lambda(t)=&&-i \omega_\lambda \alpha (t)+
\frac{kf(t)}{\sqrt{2\hbar}}\frac{v_\lambda}{\eta_+(\omega_\lambda)} 
\end{eqnarray}
which admits the solution:
\begin{equation}
\alpha_\lambda(t)=\alpha_\lambda e^{-i\omega_\lambda t}
-i \frac{k v_\lambda}{\sqrt{2\hbar}\omega_\lambda \ \eta_+(\omega_\lambda)}
\left(f(t)-\int_0^t e^{i\omega_\lambda (s-t)} \dot{f}(s)ds
\right)
\end{equation}
And we obtain the coordinate $q(t)$ in the Heisenberg picture as
\begin{eqnarray}
q(t)=i &&\sqrt{\frac{\hbar}{2}}\int d\lambda\left(
\frac{v_\lambda e^{-i\omega_\lambda t}}{\eta_-(\omega_\lambda)}\alpha_\lambda-
\frac{v_\lambda e^{i\omega_\lambda t}}{\eta_+(\omega_\lambda)}\alpha_\lambda^{+}\right) 
\nonumber \\
+&&k \int d\lambda \frac{v_\lambda^2}{|\eta_+(\omega_\lambda)|^2}\int_0^t
\sin(\omega_\lambda(s-t))f(s) ds 
\end{eqnarray}
Then, by a tedious but straightforward calculation, one obtains
the work operator ${\hat \Sigma}_\tau$ 
\begin{eqnarray}
{\hat \Sigma}_\tau 
=&&-ik\beta \sqrt{\frac{\hbar}{2}}\left\{\int d\lambda \frac{v_\lambda \hat{\dot{f}}
(\omega_\lambda;t,\tau)}{\eta_-(\omega_\lambda)}\alpha_\lambda -\int d\lambda 
\frac{v_\lambda \hat{\dot{f}}^*(\omega_\lambda;t,\tau)}{\eta_+(\omega_\lambda)}
\alpha^+_\lambda\right\} \nonumber \\ 
&&+ \int d\lambda \frac{\beta k^2 v_\lambda^2}{\omega_\lambda \mid\eta_+(\omega_\lambda)\mid^2}
\int_t^{t+\tau} ds \int_0^s ds' \cos(\omega_\lambda (s-s'))\dot{f}(s) \dot{f}(s') 
\end{eqnarray}
where $\hat{\dot{f}}(\omega_\lambda;t,\tau)$ is 
a `partial Fourier transformation' of $\dot f(t)$:
\begin{equation}
\hat{\dot{f}}(\omega_\lambda;t,\tau) \equiv \int_t^{t+\tau}\dot{f}(s)
e^{-i\omega_\lambda s} ds \ .
\end{equation} 

\subsection{Calculation of the work distribution $\pi_t$}

The distribution function $\pi_t(\Sigma_\tau=A)$ of the work 
${\hat \Sigma}_\tau$ is obtained with the aid of the characteristic
function
\begin{equation}
\Phi_t(\xi) = \int dA \ \pi_t(\Sigma_\tau=A) \ e^{i\xi A}
= \left\langle \exp\left(i\xi {\hat \Sigma}_\tau
\right)\right\rangle \ .
\end{equation}
With the aid of the formula
\begin{equation}
\langle e^{\int d\lambda(\xi_\lambda\alpha_\lambda+\eta_\lambda\alpha_\lambda^+)}\rangle
=e^{\int d\lambda\frac{\xi_\lambda \eta_\lambda}{2}
\rm{coth}(\frac{\beta \hbar \omega_\lambda}{2})}
\end{equation}
(its proof is given in Appendix), one obtains the following expression:
\begin{equation}
\Phi_t(\xi)=e^{i \xi m_{t,\tau}-\frac{\xi^2}{2}\sigma_{t,\tau}^2} \ ,
\label{ChFt}
\end{equation}
where 
\begin{equation}
m_{t,\tau}=\beta k^2 \int_t^{t+\tau} ds \int_0^s ds' 
\int d\lambda\frac{v_\lambda^2}{\mid\eta_+(\omega_\lambda)\mid^2}
\frac{\cos(\omega_\lambda(s-s'))}{\omega_\lambda}\dot{f}(s)\dot{f}(s')
\label{Mean}
\end{equation}
and
\begin{equation}
\sigma_{t,\tau}^2=\frac{\hbar k^2\beta^2}{2}\int_t^{t+\tau} ds 
\int_t^{t+\tau} ds' \int d\lambda \frac{v_\lambda^2 \cos(\omega_\lambda(s-s'))}
{\mid\eta_+(\omega_\lambda)\mid^2}{\rm coth}(\frac{\beta \hbar \omega_\lambda}{2})
\dot{f}(s)\dot{f}(s') \ .
\label{StDev}
\end{equation}
Eq.(\ref{ChFt}) shows that the distribution $\pi_t$ is Gaussian with the mean value
$m_{t,\tau}$ and the standard deviation $\sigma_{t,\tau}$:
\begin{equation}
\pi_t(\Sigma_\tau=A) = {1\over \sqrt{2\pi}\sigma_{t,\tau}}\exp\left(
-{(A-m_{t,\tau})^2\over 2\sigma_{t,\tau}^2}\right) \ . \label{Gauss}
\end{equation}

By setting $t=0$, (\ref{Gauss}) leads to the quantum counterpart of 
TFT\cite{EvansSearles0,Evans}
\begin{equation}
\log {\pi_0(\Sigma_\tau=A) \over \pi_0(\Sigma_\tau=-A)}= {2m_{0,\tau}\over
\sigma_{0,\tau}^2} A \ . \label{Q-TFT}
\end{equation}

In addition, provided that the limits $m_{+\infty,\tau}=\lim_{t\to +\infty} m_{t,\tau}$ and
$\sigma_{+\infty,\tau}=\lim_{t\to +\infty}\sigma_{t,\tau}$ exist, one has
\begin{equation}
\pi_{+\infty}(\Sigma_\tau=A) = 
{1\over \sqrt{2\pi}\sigma_{+\infty,\tau}}\exp\left(
-{(A-m_{+\infty,\tau})^2\over 2\sigma_{+\infty,\tau}^2}\right) \ . \label{Gauss2}
\end{equation}
where $\pi_{+\infty}(\Sigma_\tau=A) =\lim_{t\to +\infty}\pi_t(\Sigma_\tau=A)$,
and the quantum counterpart of SSFT is derived
\begin{equation}
\lim_{\tau\to +\infty}{1\over \tau} \log{\pi_{+\infty}(\Sigma_\tau/\tau=a) 
\over \pi_{+\infty}(\Sigma_\tau/\tau=-a)}= \lim_{\tau\to +\infty}
{2m_{+\infty,\tau}\over \sigma_{+\infty,\tau}} a \ . \label{Q-SSFT}
\end{equation}

If the coefficient of $A$ in (\ref{Q-TFT}) and that of $a$ in (\ref{Q-SSFT}) were
unity, the fluctuation theorems would hold.
However, the expressions of the mean value $m_{t,\tau}$ and
the standard deviation $\sigma_{t,\tau}$ seem to indicate the deviations.
In the following, we investigate this question for
TFT and SSFT
separately. In the latter case, since the existence of the $t\to +\infty$ 
limits strongly depends on the behavior of $f(t)$, we focus on two concrete
examples: the uniformly dragged case $f(t)=v_0 t$ and the periodically 
driven case $f(t)=v_0/\Omega \sin \Omega t$.

\section{QUANTUM CORRECTION FOR TFT}

In this section, quantum corrections for TFT (transient fluctuation theorem) will 
be investigated.

First we note that $\sigma_{0,\tau}^2$ can be rewritten as
\begin{equation}
\sigma_{0,\tau}^2={\hbar k^2\beta^2}\int_0^{\tau} ds 
\int_0^{s} ds' \int d\lambda \frac{v_\lambda^2 \cos(\omega_\lambda(s-s'))}
{\mid\eta_+(\omega_\lambda)\mid^2}{\rm coth}(\frac{\beta \hbar \omega_\lambda}{2})
\dot{f}(s)\dot{f}(s') \ .
\label{StDev2}
\end{equation}
As the integrand is different from that of $2m_{0,\tau}$, 
one has $\sigma_{0,\tau}^2\not=2m_{0,\tau}$ and, thus, TFT does not hold.

Using the expansion ${\rm{coth}}(\frac{\beta \hbar \omega_\lambda}{2})
=\frac{2}{\beta \hbar \omega_\lambda}
+\frac{1}{6}\beta \hbar \omega_\lambda+O(\hbar^3)$, 
the quantum correction can be easily calculated. Indeed, one has
\begin{eqnarray}
{\sigma_{0,\tau}}^2=&&\hbar k^2 \beta^2 \int_0^{\tau} ds 
\int_0^s ds' \int d\lambda \frac{v_\lambda^2}{\mid\eta_+(\omega_\lambda)\mid^2}
\frac{\cos(\omega_\lambda(s-s'))}{\omega_\lambda}\dot{f}(s)\dot{f}(s') \nonumber \\
+&&\frac{1}{12}\beta^3\hbar^2 k^2 \int_0^{\tau} ds \int_0^s ds' \int d\lambda 
\frac{\omega_\lambda v_\lambda^2}{\mid\eta_+(\omega_\lambda)\mid^2}
\cos(\omega_\lambda (s-s'))\dot{f}(s)\dot{f}(s') \nonumber \\
+&& O(\hbar^4)
\end{eqnarray}
and, thus,
\begin{equation}
\log {\pi_0(\Sigma_\tau=A) \over \pi_0(\Sigma_\tau=-A)}= 
\left(1+\hbar^2 \epsilon_2 +{\rm O}(\hbar^4)\right) A \ . \label{Q-TFT2}
\end{equation}
where 
\begin{equation}
\epsilon_2=-\frac{\beta^3 k^2}{24 m_{0,\tau}} \int_0^{\tau} ds \int_0^s ds' \int d\lambda 
\frac{\omega_\lambda v_\lambda^2}{\mid\eta_+(\omega_\lambda)\mid^2}
\cos(\omega_\lambda (s-s'))\dot{f}(s)\dot{f}(s') \ .
\end{equation}

To illustrate the behavior of $\epsilon_2$, we consider the uniformly dragged case
$f(t)=v_0 t$ where the damping is very weak: $\mu \equiv \int \kappa_\lambda^2 
d\lambda/k \ll 1$. Up to the first order in $\mu$, the function $1/\eta_{\pm}(\omega)$ 
can be approximated by
\begin{eqnarray}
{1\over \eta_{\pm}(\omega)} &\simeq& {1\over m(\omega-{\tilde \Omega}_0 
\pm i\gamma)(\omega+{\tilde \Omega}_0 
\pm i\gamma)} \\ \nonumber \\
{\tilde \Omega}_0&=&\sqrt{k\over m}+\int {d\lambda \over 4m}\kappa_\lambda^2
\left\{{\rm P}{1\over \sqrt{k/m}-\omega_\lambda}
+ {1\over \sqrt{k/m}+\omega_\lambda}\right\} \\
\gamma&=&{\pi\over 4m}\int {d\lambda}\kappa_\lambda^2\delta(\sqrt{k/m}-\omega_\lambda)
\end{eqnarray}
where the symbol P in the integrand stands for Cauchy's principal part.
Then, by a straightforward calculation, one finds that the leading term of the
correction $\epsilon_2$ with respect to $\mu$ is independent of time $\tau$
and 
\begin{equation}
\epsilon_2=-{\beta^2 k \over 24 m}(1+{\rm O}(\mu)) \ .
\end{equation}

\section{QUANTUM CORRECTION FOR SSFT}

In this section, quantum corrections for SSFT (steady-state fluctuation theorem) 
will be investigated.
In this case, the long-time limit $t\to +\infty$ should be taken. As the existence
and the limiting value strongly depends on the potential motion $f(t)$, we
examine quantum corrections for two examples:
the uniformly dragged case $f(t)=v_0 t$ and the periodically 
driven case $f(t)=v_0/\Omega \sin \Omega t$.

\subsection{Uniformly dragged case}

In this case $f(t)=v_0 t$ and the mean work $m_{t,\tau}$ and its variance
$\sigma_{t,\tau}^2$ read as
\begin{equation}
m_{t,\tau}=2\beta k^2 v_0^2 \int d\lambda\frac{v_\lambda^2}{\omega_\lambda 
\mid\eta_+(\omega_\lambda)\mid^2}
{\sin\left(\omega_\lambda(t+\tau/2)\right) \sin \omega_\lambda \tau/2
\over \omega_\lambda^2}
\label{MeanUni}
\end{equation}
and
\begin{equation}
\sigma_{t,\tau}^2=2\hbar k^2\beta^2 v_0^2
\int d\lambda \frac{v_\lambda^2}
{\mid\eta_+(\omega_\lambda)\mid^2}{\rm coth}(\frac{\beta \hbar \omega_\lambda}{2})
{\sin^2 \omega_\lambda \tau/2 \over \omega_\lambda^2}\ .
\label{StDevUni}
\end{equation}

In the limit of $t\to +\infty$, one finds
\begin{equation}
m_{+\infty,\tau} =\lim_{t\to +\infty} m_{t,\tau}= \int 
d\lambda\frac{2\pi\beta k^2 v_0^2 \kappa_\lambda^2}{\mid\eta_+(\omega_\lambda)\mid^2}
{\sin \omega_\lambda \tau/2
\over \omega_\lambda} \delta(\omega_\lambda) = M_{+\infty} \tau 
\label{MeanUniLim}
\end{equation}
where $v_\lambda = \sqrt{\omega_\lambda}\kappa_\lambda$ is used and
\begin{equation}
M_{+\infty} 
= {\pi\beta k^2 v_0^2 \over \mid\eta_+(0)\mid^2} 
\int d\lambda {\kappa_\lambda^2 \delta(\omega_\lambda)} 
\label{MeanUniConst}
\end{equation}
Note that $\sigma_{+\infty,\tau}^2= \sigma_{t,\tau}^2$ as the variance does not 
depend on $t$. 

These equations together with $\{\sin^2 \omega \tau/2\}/(\omega^2 \tau)\to 
\pi \delta(\omega)/2$ ($\tau\to +\infty$) give
\begin{eqnarray}
\lim_{\tau\to +\infty}{\sigma_{+\infty,\tau}^2\over 2 m_{+\infty,\tau}}&=& 
{\pi \hbar k^2\beta^2 v_0^2 \over 2 M_{+\infty}}
\int d\lambda \frac{\kappa_\lambda^2}{\mid\eta_+(\omega_\lambda)\mid^2}
\ \omega_\lambda \ {\rm coth}(\frac{\beta \hbar \omega_\lambda}{2})
\delta(\omega_\lambda) \nonumber \\
&=& {\pi \hbar k^2\beta^2 v_0^2 \over 2 M_{+\infty}}{2\over \beta \hbar
{\mid\eta_+(0)\mid^2}}
\int d\lambda \kappa_\lambda^2
\delta(\omega_\lambda) =1 \ .
\end{eqnarray}
Therefore, SSFT holds exactly:
\begin{equation}
\lim_{\tau\to +\infty}{1\over \tau} \log{\pi_{+\infty}(\Sigma_\tau/\tau=a) 
\over \pi_{+\infty}(\Sigma_\tau/\tau=-a)}= a \ . \label{Q-SSFTUni}
\end{equation}
The reason can be understood easily. Indeed, the motion of the potential 
is uniform in this case and only the zero frequency environmental modes 
do contribute to the steady-state work distribution. On the other hand,
quantum effects appear only for the non-zero frequency modes and, thus,
the quantum correction disappears in SSFT.
In the next section, we consider a case where non-zero environmental modes do
contribute to SSFT.

\subsection{Periodically driven case}

In this section, we consider the periodically driven case ${\dot f}(t)=v_0 \cos(\Omega t)$.
First we remark that, in this case, there exists no exact steady state since the 
system is periodically driven. We consider, instead, a limiting oscillatory state
obtained by a `stroboscopic limit', where the time $t$ is set to $t=2\pi m/\Omega 
+\phi/\Omega$ and the limit $m\to +\infty$ is taken. The variable $\phi$ represents
the phase of the limiting oscillation. Hereafter, we abbreviate
\begin{equation}
\lim_{m\to +\infty} F(t)|_{t=2\pi {m\over \Omega} 
+{\phi\over \Omega}} = {\rm strob.}\lim_{t\to +\infty} F(t)
\label{StrobLim}
\end{equation}

The variance of the work for $t=2\pi m/\Omega +\phi/\Omega$ now reads as
\begin{eqnarray}
\sigma_{t,\tau}^2&=&\frac{\hbar k^2\beta^2v_0^2}{2}\int_0^{\tau} ds 
\int_0^{\tau} ds' \int d\lambda \frac{v_\lambda^2 \cos(\omega_\lambda(s-s'))}
{\mid\eta_+(\omega_\lambda)\mid^2} \nonumber \\
&&\mskip 100mu \times{\rm coth}(\frac{\beta \hbar \omega_\lambda}{2})
\cos(\Omega s+\phi)\cos(\Omega s'+\phi) \ .
\label{StDevPO}
\end{eqnarray}
Since it does not depend on $m$, one has $
\sigma_{+\infty,\tau}^2\equiv {\rm strob.}\lim_{t\to +\infty}\sigma_{t,\tau}^2
= \sigma_{t,\tau}^2$.
Then, by a tedious but straightforward calculation, one obtains
\begin{equation}
\lim_{\tau\to \infty}\sigma_{+\infty,\tau}^2/\tau = {\pi\hbar k^2 \beta^2 v_0^2
\over 4}
{{\rm coth}{\beta\hbar \Omega\over 2}\over |\eta_+(\Omega)|^2}
\int d\lambda{v_\lambda^2 \delta(\omega_\lambda-\Omega)} \ .
\label{StDevLimPO}
\end{equation}

As in the same way, one has
\begin{eqnarray}
m_{+\infty,\tau}&=&{\rm strob.}\lim_{t\to +\infty}m_{t,\tau} \nonumber \\
&=&
\beta k^2 v_0^2 \int_0^{\tau} ds \int_{-\infty}^s ds' 
\int d\lambda\frac{v_\lambda^2 \cos(\omega_\lambda(s-s'))}
{\omega_\lambda\mid\eta_+(\omega_\lambda)\mid^2}
\cos(\Omega s+\phi)\cos(\Omega s'+\phi) \nonumber \\
&=&{\pi \beta k^2 v_0^2\over 4}{\tau \over \Omega |\eta_+(\Omega)|^2}
\int d\lambda v_\lambda^2 \delta(\omega_\lambda-\Omega) \nonumber \\
&&+{\beta k^2 v_0^2\over 4}\int d\lambda 
{v_\lambda^2 \over \omega_\lambda |\eta_+(\omega_\lambda)|^2} 
\biggl[{\rm P}{\cos 2\phi - \cos(2\Omega \tau+2\phi)\over 
\Omega^2-\omega_\lambda^2} \nonumber \\
&&\mskip 160 mu +\pi \delta(\Omega-\omega_\lambda)
{\sin(2\Omega \tau+2\phi)-\sin 2\phi \over 
2 \Omega} \biggr]
\label{MeanPO}
\end{eqnarray}
and 
\begin{eqnarray}
\lim_{\tau\to +\infty}m_{+\infty,\tau}/\tau
={\pi \beta k^2 v_0^2 \over 4 \Omega |\eta_+(\Omega)|^2}
\int d\lambda v_\lambda^2 \delta(\omega_\lambda-\Omega) \ .
\label{MeanLimPO}
\end{eqnarray}

Eqs.(\ref{StDevLimPO}) and (\ref{MeanLimPO}) give
\begin{equation}
\lim_{\tau\to+\infty}{2m_{+\infty,\tau}\over \sigma_{+\infty,\tau}^2}
= {2\over \beta \hbar \Omega}{\rm tanh}{\beta \hbar \Omega\over 2}
\end{equation}
and, thus, SSFT does not hold
\begin{equation}
\lim_{\tau\to +\infty}{1\over \tau} \log{\pi_{+\infty}(\Sigma_\tau/\tau=a) 
\over \pi_{+\infty}(\Sigma_\tau/\tau=-a)}= {2\over \beta \hbar \Omega}
{\rm tanh}{\beta \hbar \Omega\over 2} \ a \not= a \ . \label{Q-SSFPO}
\end{equation}
where the distribution function refers to the limiting oscillatory state.
Note that the quantum correction is universal in the sense that it depend 
only on the temperature and the frequency of the driving, but not on the 
specific features of the system and environment.

\section{DISCUSSIONS}

In summary, for a harmonic oscillator linearly coupled with a 
harmonic reservoir, we have investigated quantum analogues of 
the transient and steady-state fluctuation theorems with respect to 
the work done externally. And we have shown the followings:

\begin{itemize}

\item[(i)] In the transient fluctuation theorem, there appears 
a quantum correction of order $\hbar^2$.

\item[(ii)] In the steady-state fluctuation theorem, the existence 
of a quantum correction depends on the way of driving. In the 
uniformly dragged case, the classical formula holds, while, in the
periodically driven case, there appears a correction of order
$\hbar^2$.

\end{itemize}

\noindent
We remind that the steady-state fluctuation theorem for the periodically
driven case refers to the limiting oscillatory state, which is obtained
as the stroboscopic limit defined in (\ref{StrobLim}).

Remarkably, the situation is in contrast to that for Kurchan's 
quantum fluctuation theorem, where energy is measured twice, 
and the entropy change and work done are evaluated from the difference 
between the two measured values. In this case, as discussed 
previously\cite{9,HalTasaki,Callens,TasakiMatsui}, the (transient) fluctuation 
theorem holds as in the classical systems. 

Finally, we note that the quantum correction for
the steady-state fluctuation theorem is universal in the sense that
it depends only on the temperature and the frequency of the driving, 
but not on the specific features of the system and environment.
We expect that this is also the case for general systems.

\section*{Acknowledgement}
The authors are grateful to Professors T. Matsui and M. Sano for
fruitful discussions and valuable comments.
This work is supported by a Grant-in-Aid for
Scientific Research (C) from JSPS, by a Grant-in-Aid for Scientific 
Research of Priority Areas ``Control of Molecules in Intense Laser Fields'' 
and 21st Century COE Program (Holistic research and Education Center for
Physics of Self-Organization Systems)
both from the Ministry of Education, Culture, Sports, Science and 
Technology of Japan.

\appendix

\section{Derivation of $\langle e^{\int d\lambda(\xi_\lambda\alpha_\lambda+\eta_\lambda
\alpha_\lambda^+)}\rangle
=e^{\int d\lambda\frac{\xi_\lambda \eta_\lambda}{2}\rm{coth}(\frac{\beta \hbar \omega_\lambda}{2})}$.}

\noindent The desired result immediately follows from
\begin{equation}
\langle e^{\eta a^++\xi a}\rangle=e^{\frac{\eta\xi}{2} \rm{coth}(\frac{\beta\hbar\omega}{2})}
\ , \label{1Deg}
\end{equation}
where $[a,a^+]=1$ and $\langle \cdots \rangle\equiv {\rm tr}(\cdots e^{-\beta \hbar \omega a^+ a})/Z$
with $Z={\rm tr}(e^{-\beta \hbar \omega a^+ a})$.
This relation can be shown as follows. 

The Baker-Hausdorff theorem gives 
\begin{equation}
e^{\eta a^++\xi a}=e^{\eta a^+}e^{\xi a}e^{\frac{\xi \eta}{2}} \ .
\end{equation}
With the aid of the coherent states  $|\alpha\rangle\equiv \sum_{n=0}^\infty {\alpha^n
\over \sqrt{n!}}|n\rangle$, which satisfy
\begin{eqnarray}
a \mid \alpha \rangle = \alpha |\alpha \rangle \ , \ \ \ \ \ \langle \alpha | a^+ &=& \alpha^* 
\langle \alpha| \\
\int \frac{d\alpha}{\pi} \mid\alpha \rangle \langle \alpha\mid e^{-\mid \alpha\mid^2}=1 \ , \ \ \
\langle \alpha\mid\beta \rangle&=&e^{\alpha^* \beta} \ ,
\end{eqnarray}
one finds
\begin{eqnarray}
&&\langle e^{\eta a^++\xi a}\rangle =e^{\frac{\xi \eta}{2}} \langle e^{\eta a^+}e^{\xi a}\rangle 
=\frac{e^{\frac{\xi\eta}{2}}}{Z}{\rm tr}(e^{-\beta \hbar \omega a^+ a} 
e^{\eta a^+} e^{\xi a}) \nonumber \\
&&=\frac{e^{\frac{\xi\eta}{2}}}{Z}\sum_{n=0}^{\infty} 
e^{-\beta\hbar\omega n} \langle n\mid\int\frac{d\alpha}{\pi}\mid\alpha\rangle 
\langle \alpha\mid e^{\eta a^+}e^{\xi a} \int\frac{d\beta '}{\pi}\mid\beta'\rangle 
\langle \beta'\mid n\rangle e^{-\mid \alpha \mid^2-\mid \beta' \mid^2} \nonumber \\
&&=\frac{e^{\frac{\xi\eta}{2}}}{Z}\sum_{n=0}^{\infty} e^{-\beta\hbar\omega n} 
\int\frac{d\alpha}{\pi}\int\frac{d\beta '}{\pi}e^{-\mid \alpha \mid^2-\mid \beta' \mid^2}
\langle \alpha\mid\beta'\rangle e^{\eta \alpha^*}e^{\xi \beta'}\frac{\alpha^n{{\beta'}^*}^n}{n!} 
\nonumber \\
&&=\frac{e^{\frac{\xi\eta}{2}}}{Z}\int\frac{d\alpha}{\pi}
\int\frac{d\beta'}{\pi}e^{-\mid \alpha \mid^2-\mid \beta' \mid^2+\xi\beta'+\eta\alpha^* 
+\alpha {\beta'}^* e^{-\beta \hbar\omega}+\alpha^*\beta'} \nonumber \\
&&=e^{\frac{\xi\eta}{2}}\frac{1}{Z} \int\frac{d\beta'}{\pi}e^{-(1-e^{-\beta\hbar\omega})\mid 
\beta'\mid^2+e^{-\beta\hbar\omega}{\beta'}^*\eta+\beta'\xi}\nonumber \\
&&=e^{\frac{\eta\xi}{2} \rm{coth}(\frac{\beta\hbar\omega}{2})} \ ,
\end{eqnarray}
which leads to the desired relation (\ref{1Deg}):
\begin{equation}
<e^{\eta a^++\xi a}>=e^{\frac{\eta\xi}{2} \rm{coth}(\frac{\beta\hbar\omega}{2})}
\end{equation}

\section{Integral containing $\eta_+(\omega_\lambda)$}

To proceed the calculations in Sec.\ref{Sec.Dis}, one has to evaluate
integrals involving $\eta_+(\omega_\lambda)$. In this appendix, we
illustrate the way by evaluating
$$
\int d\lambda \frac{v_\lambda^2}{\omega_\lambda \mid\eta_+(\omega_\lambda)\mid^2} \ .
$$
The key idea is to rewrite it as a contour integral and to use the relations:
\begin{eqnarray}
\eta_-(x)-\eta_+(x)&=&-\pi i\int d\lambda \delta(x-\omega_\lambda) v_\lambda^2 \ , \\
\eta_\pm(-x)&=&\eta_\mp(x) \ .
\end{eqnarray}
Then, one has
\begin{eqnarray}
&&\int d\lambda \frac{v_\lambda^2}{\omega_\lambda \mid\eta_+(\omega_\lambda)\mid^2} 
=\int d\lambda \int_{0}^{\infty} dx \delta(x-\omega_\lambda)\frac{v_\lambda^2}{\eta_-(x)\eta_+(x)}\frac{1}{x} \nonumber \\
=&&\int_{0}^{\infty}\frac{1}{-\pi i}(\frac{1}{\eta_+(x)}-\frac{1}{\eta_-(x)})\frac{1}{x}dx 
=\int_{-\infty}^{\infty}\frac{1}{-\pi i}\frac{1}{\eta_+(x)}\frac{1}{x}dz \nonumber \\
=&&\int_{C}\frac{1}{-\pi i}\frac{dz}{z \eta(z)} 
-\lim_{r\rightarrow  0}\int_0^{2 \pi}  d\theta \frac{1}{-\pi i}
\frac{r i e^{i \theta}}{r e^{i \theta} \eta(r e^{i \theta})} 
-\lim_{R\rightarrow \infty}\int_0^{2 \pi} \frac{1}{-\pi i}
\frac{R i  e^{i \theta} d\theta }{R e^{i \theta} \eta(R e^{i \theta})} \nonumber \\
=&&\frac{1}{k} \ , \label{CalC}
\end{eqnarray} 
where the integration contour $C$ is shown in Fig.B.1.

\begin{figure}
\center{
\includegraphics[scale=0.8]{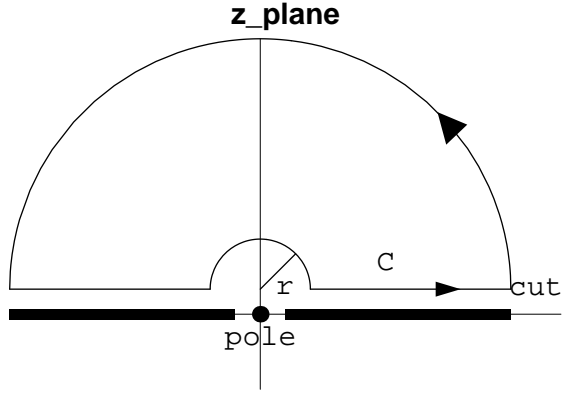}
}
\caption{Integration contour for the integral (\ref{CalC})}
\label{Figg}
\end{figure}

\end{document}